\documentclass[aps,prl,twocolumn,secnumarabic,amssymb, amsmath,nobibnotes,showpacs]{revtex4-1}

\usepackage{amssymb,amsmath,amscd}
\usepackage{graphicx,calc,epsfig,pstricks,bbm}
\usepackage{tikz}
\usepackage{enumerate}
\expandafter\ifx\csname package@font\endcsname\relax\else
 \expandafter\expandafter
 \expandafter\usepackage
 \expandafter\expandafter
 \expandafter{\csname package@font\endcsname}%
\fi

\newcommand {\be}{\begin{equation}}
\newcommand {\ee}{\end{equation}}
\newcommand{\ba}{\begin{array}{c}}
\newcommand{\ea}{\end{array}}

\newcommand{\cube}{\ba
 \begin{tikzpicture}
\pgfmathsetmacro{\cubex}{0.15}
\pgfmathsetmacro{\cubey}{0.15}
\pgfmathsetmacro{\cubez}{0.15}
\draw (0,0,0) -- ++(-\cubex,0,0) -- ++(0,-\cubey,0) -- ++(\cubex,0,0) -- cycle;
\draw (0,0,0) -- ++(0,0,-\cubez) -- ++(0,-\cubey,0) -- ++(0,0,\cubez) -- cycle;
\draw (0,0,0) -- ++(-\cubex,0,0) -- ++(0,0,-\cubez) -- ++(\cubex,0,0) -- cycle;
\end{tikzpicture}
\ea}
\newcommand{\vgraph}{\mathfrak{n}}

\begin{document}
\title{Loop Quantum Cosmology from Loop Quantum Gravity}%

\author{Emanuele Alesci$^{1}$, Francesco Cianfrani$^{2}$}%
\affiliation{$^{1}$Instytut Fizyki Teoretycznej, Uniwersytet Warszawski, ul. Pasteura 5, 02-093 Warszawa, Poland, EU\\
$^{2}$Institute
for Theoretical Physics, University of Wroc\l{}aw, Pl.\ Maksa Borna
9, Pl--50-204 Wroc\l{}aw, Poland, EU.}
\date{\today}%

\begin{abstract} 
\noindent 
We show how Loop Quantum Cosmology can be derived as an effective semiclassical description of Loop Quantum Gravity. Using the tools of QRLG, a gauge fixed version of LQG, we take the coherent states of the fundamental microscopic theory suitable to describe a Bianchi I Universe and we find a mapping between the expectation value of the Hamiltonian and the dynamics of LQC. Our results are in agreement with a lattice refinement framework for LQC, thus the so called ``old'' and ``improved-dynamics'' regularization schemes can be reproduced. These amount to different choices of relations between local variables and the smeared ones entering the definition of the coherent states. The leading order of the fundamental theory corresponds to LQC, but we also find different inverse volume corrections, that depend on a purely quantum observable, namely the number of nodes of the states.
\end{abstract}


\maketitle

The attempts towards a quantum theory for the gravitational field encountered several interpretative and conceptual issues. The latter are partially tamed in a cosmological setting, in which some degrees of freedom are frozen and the dynamic problem simplifies. This is the case already for the Wheeler-DeWitt framework, whose implementation in minisuperspace \cite{Misner:1969ae} allows to go further in the quantization procedure, even though no new achievement in the characterization of a Quantum Universe is obtained (no removal of singularity or of chaotic behavior for Bianchi IX model). On the contrary, in Loop Quantum Cosmology (LQC) \cite{Bojowald:2011zzb,Ashtekar:2011ni,Banerjee:2011qu} a viable description for the early Universe is provided: the initial singularity is generically replaced by a Big-Bounce \cite{Ashtekar:2006rx}, while there are some indications that the chaotic structure of the Bianchi IX solution is tamed \cite{Bojowald:2003xe}. 

The investigation of the phenomenological implications of LQC \cite{Bojowald:2011hd,Agullo:2012sh} (see \cite{Calcagni:2012vw} for a review) opens up the possibility to validate the theory in a nearby future in view of the continuous progress of experimental cosmology (CMB power spectrum and polarization, neutrino cosmology, gravitational waves detection). This validation would not only give us a detailed knowledge of the Universe early phases, but it could also enhance our comprehension of fundamental fields at the Planck scale, given the expected link between LQC and Loop Quantum Gravity (LQG) \cite{revloop}. In fact, LQC aims to provide the cosmological sector of LQG, but till now no direct derivation of the former from the latter has been given.   

In this work, we bridge the two theories. The intermediate step we use is Quantum-Reduced-Loop-Gravity (QRLG) \cite{Alesci:2012md,Alesci:2013xd,Alesci:2013xya}, which is realized by imposing some gauge-fixing conditions weakly in the kinematical Hilbert space of LQG. These gauge-fixing conditions restrict the full Hilbert space to one describing a diagonal metric tensor and diagonal (inverse) drei-bein fields. This formulation has the merit to preserve a (cuboidal) graph structure with SU(2) quantum numbers at links and intertwiners at nodes, thus accounting for the main features of LQG in a simplified context. 

The dynamics of the resulting system is analyzed out of the one of the full theory by replacing LQG operators with QRLG ones. This actually amounts to reducing the degrees of freedom (since generically the dynamics does not preserve the diagonal condition for the metric) to the ones of a inhomogeneous extension of the Bianchi I model. The action of the scalar constraint operator can then be analytically described. The semiclassical limit has been realized by peaking coherent states around a classical configuration: for a three-valent node it has been demonstrated that the expectation values of the scalar constraint coincides with the expression adopted in LQC for the Bianchi I model \cite{Alesci:2014uha}. 

This result has the following drawbacks, which we are going to solve in this work: 
\begin{enumerate}[i)]
\item  it was restricted to a single node with ad-hoc structure, namely only inside an homogenous patch of the inhomogeneous Bianchi I global state;

\item  it was not clear the relationship between the local regulators in QRLG and the global ones in LQC.

\end{enumerate}
In particular, we are going to outline how to refine our calculations such that 
the extension to more-than-three valence nodes in an arbitrary cuboidal graph results to be straightforward. Then, we will discuss the role of the regulator in QRLG, which is a  parameter entering the definition of the classical phase space variables used to peak the semiclassical states, and we will demonstrate  how it can be matched with the regulator adopted in LQC, in which it enters the definition of quantum states. In particular, we will outline how the prescription of lattice refinement \cite{Bojowald:2006qu,Bojowald:2011iq} is naturally inferred, such that different regularization prescriptions in LQC (as the ``old and ``improved dynamics'' schemes) are related with the different definitions one can give to the classical local variables (the scale factors) from smeared ones (fluxes).
 
These findings fix the correspondence between the dynamics of LQC and the effective semiclassical description of QRLG and sustain the idea that the former correctly account for the fundamental structure of a quantum Universe in LQG. 

However, we find that some (inverse volume) corrections are enhanced compared to the analogous ones in LQC by a factor equal to the square root of the total number nodes. This feature gives more chances to validate the model through the comparison with experimental data.   

\paragraph{Loop Quantum Cosmology} In LQC, a flat homogeneous Universe is described by the conjugate variables $\{c_i,p^j\}$ giving the extrinsic curvature component and the scale factor. By using some tools of LQG, a inequivalent scenario with respect to Wheeler-DeWitt minisuperspace quantization is obtained. In particular, wave function are quasi-periodic functions of $c_i$ and the Hilbert space is the direct product of three Bohr compactifications of the real line. The gravitational dynamics is described by the scalar constraint, whose expression reads \cite{MartinBenito:2008wx,Ashtekar:2009vc} (modulo a factor ordering)
\begin{align}
H=\frac{2}{\gamma^2}\mathcal{N} \bigg(&\sqrt{\frac{p^xp^y}{p^z}}\,\frac{\sin{\mu_x c_x}}{\mu_x}
\,\frac{\sin{\mu_y c_y}}{\mu_y}+\nonumber\\
&\sqrt{\frac{p^yp^z}{p^x}}\,\frac{\sin{\mu_y c_y}}{\mu_y}
\,\frac{\sin{\mu_z c_z}}{\mu_z}+\nonumber\\
&\sqrt{\frac{p^zp^x}{p^y}}\,\frac{\sin{\mu_z c_z}}{\mu_z}
\,\frac{\sin{\mu_x c_x}}{\mu_x}\bigg)\,,\label{Hlqc}
\end{align}
$\gamma$ and $\mathcal{N}$ being the Immirzi parameter and the homogeneous lapse function, while $\mu_i$ denotes the regulators, for which two possible choices have been considered in literature: the ``old'' ($\mu_0$) scheme, $\mu_i=\mu_0=const.$, and the ``improved-dynamics'' ($\bar\mu$) scheme, in which the regulators $\mu_i=\bar\mu_i$ depend on $p$'s such that
\be
\bar\mu_x\bar\mu_y=\frac{\Delta l_P^2}{p^z}\,,\quad
\bar\mu_y\bar\mu_z=\frac{\Delta l_P^2}{p^x}\,,\quad
\bar\mu_z\bar\mu_x=\frac{\Delta l_P^2}{p^y}\,,\label{barmu}
\ee
$l_P$ being the Planck length. The factor $\Delta=4\pi\sqrt{3}\gamma$, is the minimum physical area in $l_P^2$ units, it's imported from the full theory to regularize the minimum size of the holonomies entering the definition of the quantum states. Effectively, one gets from \eqref{Hlqc} a modified Friedman equation, in which the initial singularity is replaced by a bounce occurring at a critical energy density $\rho=\rho_{cr}\propto \mu^2 p$ in the isotropic limit \cite{Singh:2006im}. Generically, in the isotropic limit the old and improved-dynamics regularization schemes are only two particular cases of the lattice refinement of LQC \cite{Bojowald:2006qu,Bojowald:2011iq}, in which the regulator goes as the inverse third root of the total number of nodes $N$ of the graph underlying the continuous space picture, {\it i.e.} 
\be\label{latref}
\mu\propto 1/N^{1/3}, 
\ee
and as a function of $p$ one gets $\mu\propto p^{-n}$, $n=0,1/2$ being the two schemes previously discussed.

The corrections with respect to the classical dynamics are of two kinds \cite{Bojowald:2007cd,Bojowald:2007hv}: holonomy corrections, which arise from the expansion of $\sin{\mu_i c_i}/\mu_i$ for $\mu_i c_i\ll 1$ and inverse volume corrections, which are due to the technical tools adopted to define the operator $1/\sqrt{p}_i$, {\it i.e.}
\begin{align}
\hat{\left(\frac{1}{\sqrt{p^i}}\right)}\equiv &\frac{\sqrt{p^i+8\pi\gamma l_P^2\mu_i}-\sqrt{p^i-8\pi\gamma l_P^2\mu_i}}{8\pi\gamma l_P^2\mu_i}=\\
&\frac{1}{\sqrt{p^i}}\left(1+\mu^2_i\, O\left(\frac{8\pi\gamma l_P^2}{p^i}\right)^2\right)\,.\label{lqcinvcorr}
\end{align}
The study of corrections is a tantalizing subject of investigation in view of constructing a phenomenology to be compared with observations of the CMB power spectrum \cite{Bojowald:2011hd}.   

\paragraph{Quantum Reduced Loop Gravity} QRLG has been realized in order to investigate the quantum properties of symmetry-reduced sectors of general relativity. This is done starting from the  full kinematical Hilbert space $\mathcal{H}^{kin}$ of LQG, whose elements are labeled by  oriented graphs $\Gamma$ in the spatial manifold and are functions on $L$-copies of $SU(2)$, $L$ being the number of links in $\Gamma$. A basis of states is obtained by assigning an irreducible representations ${\bf j_l}$ of $SU(2)$ on each link $l$, and a $SU(2)$ intertwiner $x_n$ at each node $n$, the latter implementing SU(2) gauge invariance. 

The basic idea of QRLG is to perform on a quantum level the following two gauge fixings \cite{Alesci:2013xya}:
\begin{enumerate}
\item the choice of a diagonal metric tensor, which is realized by restricting to cuboidal graphs, whose links $l_i$ are only those along some fiducial directions $\vec{u}^i$; 
\item the choice of a diagonal tetrad field, which implies breaking the group of internal SU(2) rotation, via the projection of SU(2) group elements based at $l_i$ to the U(1) ones obtained by stabilizing along the direction $\vec{u}^i$. This is done by using coherent states along the directions $\vec{u}^i$ and $-\vec{u}^i$ (which corresponds to the maximum and minimum magnetic number along the direction $\vec{u}^i$).
\end{enumerate}

Finally, the elements of the reduced kinematical Hilbert space are labeled by cuboidal graphs, having U(1) group elements at links and some reduced intertwiners at nodes. Such reduced intertwiners are not just those of the U(1) group, but, for nodes connecting links along different fiducial directions, they have to be evaluated from the full SU(2) intertwiners. A key property is that, since U(1) irreps are one-dimensional, reduced intertwiners are just complex numbers. This feature simplifies significantly the semiclassical analysis. 

In order to study the action of operators, we developed a reduced recoupling theory \cite{Alesci:2013xd},
consisting in recoupling holonomies as U(1) group elements and intertwiners as $SU(2)$ ones and then projecting on the representation with maximum magnetic numbers.

The scalar constraint has been defined by taking the expression of the Euclidean part in LQG and by substituting SU(2) operators with reduced ones \cite{Alesci:2012md}. Thanks to the fact that the volume operator is now diagonal, the matrix elements between states with three-valence nodes could be analytically evaluated.  This feature allows then to analyze the {\it local} behaviour of the fundamental quantum geometry along the same lines of LQG, namely as a quantum spin dynamics \cite{qsd}, without any reference to a coordinate dependent background structure. The semiclassical analysis \cite{Alesci:2014uha} has been performed by defining semiclassical states (according with LQG techiques \cite{Thiemann:2000bw, Thiemann:2002vj}) over a single three-valence dressed nodes, {\it i.e.} containing the loops added by the scalar constraint itself. This implies that the regulator, being associated with the area  of such loop, is a parameter labeling the semiclassical state. 

By performing a large $j$'s limit, where $j$'s denote the quantum number around which semiclassical states are peaked, the expectation value of the scalar constraint has been computed. The result is that at each node the leading term resembles the expression of the scalar constraint for the Bianchi I model in LQC \eqref{Hlqc}:
\begin{align}
\langle\, {}^{R}\hat{H}^{1/2}_{\cube}\,\rangle_{\vgraph} \,\approx
\frac{2}{\gamma^2}\mathcal{N}(\vgraph)\bigg(&\sqrt{\frac{p^x\;p^y}{p^z}}\;  \sin{c_x} \sin{c_y}+\nonumber\\
&+\sqrt{\frac{p^y\;p^z}{p^x}}\; \sin{c_y} \sin{c_z}+\nonumber\\
&+\sqrt{\frac{p^z\;p^x}{p^y}}\; \sin{c_z} \sin{c_x}\bigg).
\label{h1cell}
\end{align}  
$\mathcal{N}(\vgraph)$ being the lapse function at the considered node $\vgraph$, while
$c_i$ and $p^i$ denote the semiclassical values of physical phase-space variables (they are rescaled via the length and area of the dressing loops with respect to the analogous expression in \cite{Alesci:2014uha}). 

The spin quantum numbers are related to $p^i$ throught 
\be\label{pj}
p^i= 8\pi\gamma l_P^2\, j_i\,,
\ee
and the factors $1/\sqrt{p^i}$ inside \eqref{h1cell} come from the large $j$ expansion of the following expression
\be
\sqrt{j_i+1}-\sqrt{j_i-1}=\frac{1}{\sqrt{j_i}}\left(1+O(j_i^{-2})\right)\,.
\ee
The next-to-leading order terms give rise to inverse volume corrections
\be
\frac{1}{\sqrt{p_i}}\rightarrow \frac{1}{\sqrt{p_i}}\left[1+O\left(\frac{8\pi\gamma l_P^2}{p_i}\right)^2\right]. \label{invvol}
\ee
  
We are now going to outline how it is possible to refine our construction, 
such that the semiclassical analysis can be performed for more-than-three-valence nodes and  beyond the local patch approximation
. 
Then, we will 
 show how 
our regularization procedure leads to the $\bar\mu$ scheme adopted in improved dynamics. 

\paragraph{Reduced recoupling.}
We change the reduced recoupling theory in the following way:
we still recouple the U(1) group elements according with the U(1) composition law, while we use standard SU(2) recoupling theory for intertwiners. The reason why we used the ``old'' recoupling theory was that the composition of reduced intertwiners produced another reduced intertwiner. This is not the case anymore, but being reduced intertwiners one-dimensional, their composition can simply be written as a (Clebsh-Gordan) coefficient  times a reduced intertwiner.

 This slight change has the following implications on previous calculations \cite{nuovo}: 
\begin{enumerate}[i)]       
\item the action of the scalar constraint (equation (53) in \cite{Alesci:2014uha}) is modified through some non-trivial factors (which can be easily computed),
\item we have a better constrol of the corrections to the scalar constraint semiclassical expectation value.  
\end{enumerate}
Furthermore, the semiclassical analysis can be repeated also for higher valence nodes and the results is just the three-valence one summed over all the possible permutations of links. 
This allow us to investigate the generic case in which the nodes are six-valence and to construct a model for a quantum Universe as a collection of cubic cells. Hence, we don't need dressed nodes anymore, and we can regularize the scalar constraint by considering holonomies along fundamental plaquettes of the six-valence graph. 

The expectation value of the scalar constraint is thus just the sum over all the nodes and over all possible triples of links of the three-valence result. Peak around an homogeneous anisotropic configuration (the same expectation values for each cell) made of equal gaussians at links along the same fiducial directions, we get \eqref{h1cell} times the total number nodes $N$. The physical phase-space variables $P^i,C_i$ describing a collection of cells with $N$ nodes are obtained from those of a single-cell via the replacement
\begin{align} 
&p^x\rightarrow P^x=N_y N_z p^x,\; p^y\rightarrow P^y=N_z N_x p^y,\; p^z\rightarrow P^z=N_x N_y p^z,\label{resc}\\
&c_x\rightarrow C_x=N_x c_x\,,\quad c_y\rightarrow C_y= N_y c_y\,,\quad c_z\rightarrow C_z=N_z c_z\,, 
\end{align}
$N_i$ being the number of nodes lying in the dual plane to the fiducial direction $\vec{u}^i$ and $N=N_xN_yN_z$. 

Finally, the expectation value of the scalar constraint for a Bianchi I patch made of a collection of cells with $N$ nodes is given by
\begin{align}
\langle\, {}^{R}\hat{H}^{1/2}\,\rangle_{N} \,\approx
\frac{2}{\gamma^2}\mathcal{N}\bigg(&N_x\, N_y\,\sqrt{\frac{P^x\;P^y}{P^z}}\;  \sin{\frac{C_x}{N_x}} \sin{\frac{C_y}{N_y}}+\nonumber\\
&+N_y\, N_z\,\sqrt{\frac{P^y\;P^z}{P^x}}\;  \sin{\frac{C_y}{N_y}} \sin{\frac{C_z}{N_z}}+\nonumber\\
&+N_z\, N_x\,\sqrt{\frac{P^z\;P^x}{P^y}}\;  \sin{\frac{C_z}{N_z}} \sin{\frac{C_x}{N_x}}\bigg).
\end{align}
This expression coincides formally with the quantum operator adopted in LQC \eqref{Hlqc}, as soon as we make the identification  
\be\label{muN}
\mu_i=1/N_i. 
\ee
The $\mu$ parameter introduced as an ad-hoc function in LQC, now acquires a precise meaning representing the ratio of the patch-to global lattice size, as in lattice refinement of LQC \cite{Bojowald:2006qu,Bojowald:2011iq}. In fact, \eqref{muN} provides an extension of \eqref{latref} to the anisotropic case. 

Therefore, we get a fundamental correspondence between the effective semiclassical dynamics of QRLG and the quantum Hamiltonian adopted in LQC. This achievement is in agreement with the analogous results of \cite{Gielen:2013kla,Gielen:2013naa,Calcagni:2014tga} concerning cosmology in a Group Field Theory framework.  
 Let us note that from \eqref{pj} and \eqref{resc}, we get 
\begin{align}
\frac{1}{N_x}\frac{1}{N_y}=\frac{8\pi\gamma l_P^2}{P^z} j_z\,,&\quad
\frac{1}{N_y}\frac{1}{N_z}=\frac{8\pi\gamma l_P^2}{P^x} j_x\,,\nonumber\\
&\frac{1}{N_z}\frac{1}{N_x}=\frac{8\pi\gamma l_P^2}{P^y} j_y\,.\label{1/N}
\end{align}
Among the possible regularization prescriptions, the old scheme is reproduced for $\mu_0=1/N_i$, which implies that the number of nodes does not depend on $P_i$, thus $P^i$ is a linear function of $j_i$. Similarly, the improved-dynamics scheme is obtained once we fix $\bar\mu_i=1/N_i$, if we restrict to constant $j_i$'s (independent of $P$'s) such that \eqref{barmu} is reproduced from \eqref{1/N}. Henceforth, the $\bar\mu$ regularization scheme is obtained by fixing $j$'s and taking a $P$ dependent number of nodes. 
These relationships between the features of fundamental graphs (number of nodes and spin numbers) with the regularization schemes have been anticipated in \cite{Bojowald:2006qu}.

\paragraph{Corrections} The study of correction reveals that there is no exact matching of semiclassical QRLG and LQC. While holonomy corrections are the same, inverse volume ones differ. In fact, the latter are obtained from those of a single cell \eqref{invvol} and, after the rescaling \eqref{resc}, a factor $N/N_i$ appears in front of the next-to-leading order terms
\be
\frac{1}{\sqrt{P_i}}\rightarrow \frac{1}{\sqrt{P_i}}\left[1+\frac{N^2}{N^2_i}\,O\left(\frac{8\pi\gamma l_P^2}{P_i}\right)^2\right]. 
\ee
By comparing the expression above with that of LQC \eqref{lqcinvcorr}, we see the presence of an additional factor $N^2$, which enhance inverse volume corrections. 
It is worth noting how in the isotropic case, by assuming $\mu=1/N^{1/3}\propto P^{-n}$, inverse volume corrections are of order $N^{4/3}/P^2\propto P^{2(2n-1)}$ and they are negligible at high scale factors for $n\leq 1/2$, the upper bound given by the improved dynamics scheme. This provides us with a different range of admissible values for $n$ with respect to LQC for what concerns viable inverse volume corrections.

\paragraph{Conclusions}
We derived the scalar constraint adopted in LQC for the Bianchi I model from the semiclassical limit of the scalar constraint operator in QRLG. This achievement shows how LQC captures the local dynamics of full LQG, after the gauge fixing to diagonal metric tensor and triads and a reduction of the dynamical degrees of freedom to those of the Bianchi I model.
The effective regulator of LQC is the inverse of the number of nodes over each fiducial plane and this result clarifies how taking the continuum limit corresponds to removing the regulator.
However, the study of inverse volume corrections reveals that they are enhanced with respect to LQC by a factor proportional to the square of the total number of nodes (which becomes an observable on a quantum level \cite{Gambini:2013ooa,Gielen:2014uga}). This feature leaves more room for phenomenological implications than LQC. All these achievements outline the feasibility of QRLG to extract predictions out of LQG.

{\it Acknowledgment}
We wish to thank Tomasz Pawlowski for useful discussions.   
FC is supported by funds provided by the National Science Center under the agreement
DEC-2011/02/A/ST2/00294. The work of E.A. was supported by the grant of Polish Narodowe Centrum Nauki nr DEC-2011/02/A/ST2/00300.
This work has been partially realized in the framework of the CGW collaboration (www.cgwcollaboration.it).



%

\end{document}